\documentclass[twocolumn,showpacs,amsmath,amssymb]{revtex4}
\input{epsf}

\usepackage{graphicx}
\usepackage{array}
\usepackage{bigstrut}
\usepackage{longtable}
\usepackage{rotating,booktabs}
\usepackage{booktabs,threeparttable}
\usepackage{bm}
\usepackage{lipsum}

\begin{document}

\title{Dynamic multipolar polarizabilities and hyperpolarizabilities of the Sr lattice clock}

\author{Fang-Fei Wu,$^{1,4}$ Yong-Bo Tang,$^{2,3,*}$~\footnotetext{*Email Address: ybtang@htu.edu.cn} Ting-Yun Shi,$^{1}$ and Li-Yan Tang$^{1,\dag}$~\footnotetext{\dag Email Address: lytang@wipm.ac.cn}}

\affiliation {$^1$ State Key Laboratory of Magnetic Resonance and
Atomic and Molecular Physics, Wuhan Institute of Physics and Mathematics, Chinese Academy of Sciences, Wuhan 430071, People's Republic of China}
\affiliation {$^2$ College of Engineering Physics, Shenzhen technology University, Shenzhen 518118, China}
\affiliation {$^3$ College of Physics and Materials Science, Henan Normal University, Xinxiang 453007, People¡¯s Republic of China}
\affiliation {$^4$ University of Chinese Academy of Sciences, Beijing 100049, People's Republic of China}

\date{\today}

\begin{abstract}
The progress in optical clock with uncertainty at a level of $10^{-18}$ requires unprecedented precision in estimating the contribution of multipolar and higher-order effects of atom-field interactions. Current theoretical and experimental results of dynamic multipolar polarizabilities and hyperpolarizabilities at the magic wavelength for the Sr clock differ substantially. We develop a combined approach of the Dirac-Fock plus core polarization (DFCP) and relativistic configuration interaction (RCI) methods to calculate dynamic multipolar polarizabilities and hyperpolarizabilities of the Sr atom. Our differential dynamic hyperpolarizability at the magic wavelength is $-2.09(43)\times10^{7}$ a.u., which is consistent with the existing theoretical and experimental results. Our differential multipolar polarizability is $2.68(94)\times 10^{-5}$ a.u., which validates independently the theoretical work of Porsev {\em et al.} [Phys. Rev. Lett. 120, 063204 (2018)], but different from recent measurement of Ushijima {\em et al.} [Phys. Rev. Lett. 121, 263202 (2018)].
\end{abstract}

\pacs{31.15.ac, 31.15.ap, 34.20.Cf}
\maketitle

\section{Introduction}
The last few years have witnessed significant advances in optical clocks, which enable a wide range of applications, such as redefine the unit of time~\cite{bregolin17a,yamanaka15a}, test the local Lorentz invariance~\cite{bars17a,shaniv18a}, probe dark matter and dark energy~\cite{arvanitaki15a,roberts17a}, search variations of the fundamental constants~\cite{godun14a,huntemann14a,safronova18a}, and detect gravitational wave~\cite{kolkowitz16a}. At present, the highest fractional accuracy of optical clocks has reached the level of $10^{-19}$ based on the Al$^+$~\cite{brewer19a}, while the uncertainty for the Sr~\cite{nicholson15a,campbell17a} and Yb~\cite{mcgrew18a} optical lattice clocks has achieved an accuracy of $10^{-18}$ level. Aiming to develop optical clock with uncertainty and stability below $10^{-18}$, a better understanding and meticulous control of the atom-field interactions would benefit for the realization of a new generation of higher-precision optical clocks.

Stark shift as one significant sources of systematic uncertainty for most clocks~\cite{nicholson15a,campbell17a,mcgrew18a}, it is closely related to the polarizabilities and hyperpolarizabilities of clock states. Employing a magic wavelength optical lattice~\cite{katori99b,katori03a,ye99} can eliminate the leading-order of Stark shift, but can not cancel the residual multipolar and higher-order Stark shifts for optical lattice clocks. At the level of $10^{-19}$ accuracy, the effects on the systematic uncertainty of optical clock from the multipolar and higher-order atom-field interaction need to be quantitatively evaluated~\cite{ovsiannikov13a,katori15a,porsev18a,ushijima18a}.

For the Sr clock, the differential dynamic multipolar polarizability of $\Delta\alpha^{QM}(\omega)$ at the magic wavelength has contradictions among available theoretical and experimental results. The latest measurement is $-8.01(33)\times10^{-5}$ a.u.~\cite{ushijima18a}, which disagrees with the recent theoretical result of $2.80(36)\times10^{-5}$ a.u.~\cite{porsev18a} and previous experimental value of $0.0(2.6)\times10^{-5}$ a.u.~\cite{westergaard11a}. Especially the sign in $\Delta\alpha^{QM}(\omega)$ between the measurement~\cite{ushijima18a} and theory~\cite{porsev18a} is opposite each other. In addition, the differential dynamic hyperpolarizability at the magic wavelength also has discrepancy in theory and experiment. The recent RIKEN experimental result of $-2.10(7)\times10^7$ a.u.~\cite{ushijima18a} agrees well with the SYRTE measurement of $-2.01(45)\times10^7$ a.u.~\cite{westergaard11a,targat13a} and the theoretical calculation of $-1.5(4)\times10^7$ a.u.~\cite{porsev18a}, but it is inconsistent with the zero value of $-1.3(1.3)\times10^7$ a.u. measured by JILA~\cite{nicholson15a}. Especially, the single-electron approximated result of $-3.74\times10^7$ a.u.~\cite{katori15a} is not within the error bar of any other existing theoretical and experimental results. Therefore, carrying out an independent theoretical calculation is expected to solve these discrepancies.

In this paper, we develop an effective method by combining the Dirac-Fock plus core polarization (DFCP) and relativistic configuration interaction (RCI) approaches for the relativistic calculation of the divalent atoms, and apply it to calculate the dynamic multipolar polarizabilities and hyperpolarizabilities at the magic wavelength for the Sr clock states by employing the sum-over-states method. The detailed comparisons for the energies, reduced matrix elements and static dipole polarizabilities between our results and other published literatures are also made. Our work p not only resents an independent test for the previous calculations of Ref.\cite{porsev18a}, but also would stimulate further investigations on the differential multipolar polarizability of the Sr clock. The atomic units (a.u.) are used throughout of this work except specifically mentioned.

\section{Theoretical framework}
\subsection{The combination method of DFCP and RCI }
The basic strategy of present theoretical method is that a divalent electron atom is simplified as a frozen core part and valence electron part. The calculation process can be divided into three steps. The first step is the Dirac-Fock (DF) calculation of frozen core part to obtain the core orbital functions $\psi(\bm{r})$~\cite{tang14a}.

The second step is to solve the followed DFCP equation to obtain the single-electron wave functions $\phi(\bm{r})$,
\begin{equation}
h_{\rm DFCP}(\bm{r})\phi(\bm{r})=\varepsilon\phi(\bm{r})
\,,\label{e3}
\end{equation}
here $h_{\rm DFCP}(\bm{r})$ represents the DFCP Hamiltonian,
\begin{equation}
h_{\rm DFCP}(\bm{r})=c{{\bm{\alpha}}}\cdot{\mathbf p}+(\beta-1)c^{2}+V_{N}(r)+V_{\rm core}(r)
\,,\label{e4}
\end{equation}
where $\bm{\alpha}$ and $\beta$ are the $4\times 4$ Dirac matrices, $\mathbf p$ is the momentum operator for the valence electron, $V_{N}(r)$ is the Coulomb potential between a valence electron and nucleus, $V_{\rm core}(r)$ represents the interaction potential between core electrons and a valence electron, which is approximated as a DF potential and a semi-empirical one body core-polarization interaction potential~\cite{mitroy88c},
\begin{equation}
V_{\rm core}(r)=V_{\rm DF}(r)+V_{\rm 1}(r)
\,,\label{e5}
\end{equation}
with
\begin{eqnarray}
V_{1}(r)=-\frac{\alpha_{\rm core}}{2r^4}[1-exp(\frac{r^6}{\rho_\kappa^6})]
\,,\label{e6}
\end{eqnarray}
where $\alpha_{\rm core}=5.812$ a.u.~\cite{safronova10f} is the static dipole polarizability of the Sr$^{2+}$ core. $\rho_\kappa$ is the radial cutoff parameter which is tuned to reproduce the experimental binding energy of the lowest state of each $\kappa$ angular quantum number. The values of our cutoff parameters for different $\kappa$ are listed in Table~\ref{t1}. The core wave functions $\psi(\bm{r})$ obtained in the first step are used to evaluate the matrix elements of the DF potential $V_{\rm DF}(r)$~\cite{tang14a}.

The third step is configuration interaction calculation of a divalent electron atom. The eigen equation can be expressed as
\begin{eqnarray}
 (\sum_i^{2}h_{\rm DFCP}(\bm{r_i})+V_{ij})|\Psi({\pi}JM)\rangle=E|\Psi({\pi}JM)\rangle
\,.\label{e7}
\end{eqnarray}

The two-particle interaction potential is
\begin{eqnarray}
V_{ij}=\frac{1}{r_{ij}}+V_{2}(r_{ij})
\,,\label{e8}
\end{eqnarray}
the first term is the Coulomb interaction between two valence electrons, the second term is two-body core-polarization interaction with the functional form~\cite{mitroy10a,mitroy03f},
\begin{eqnarray}
V_{2}(r_{ij})=-\frac{\alpha_{\rm core}\bm {r}_i\cdot\bm{r}_j}{r_i^3r_j^3}\sqrt{[1-exp(\frac{r_i^6}{{\rho'_\kappa}^6})][1-exp(\frac{r_j^6}{{\rho'_\kappa}^6})]}
\,\label{e9}
\end{eqnarray}
where $\rho'_\kappa$ is fine-tuned on the $\rho_\kappa$ that optimized in the second step to get accurate energy for the divalent atoms.

The wave function $|\Psi({\pi}JM)\rangle$ with parity $\pi$, angular momentum $J$, and magnetic quantum number $M$ of the system is expanded as a linear combination of the configuration-state wave functions $|\Phi_{I}(\sigma{\pi} JM)\rangle$,  which are constructed by the single-electron wave functions $\phi(\bm{r})$ obtained in the second step~\cite{chen93a,zhang15}.
\begin{eqnarray}
|\Psi({\pi}JM)\rangle=\sum_{I}{C_{I}|\Phi_{I}(\sigma{\pi} JM)\rangle}
\,,\label{e10}
\end{eqnarray}
where $C_{I}$ and $\sigma$ are the expansion coefficients and the additional quantum number to define each configuration state uniquely, respectively. Throughout the present calculations, the basis functions are constructed by using the Notre Dame basis sets~\cite{johnson88a}.
\begin{table}[!htbp]
\caption{\label{t1}  The radial cutoff parameter $\rho_\kappa$ (in a.u.) for different quantum state.}
\begin{ruledtabular}
\begin{tabular}{llllll}
\multicolumn{1}{c}{$\kappa=-1$}&\multicolumn{1}{c}{$\kappa=1$}&\multicolumn{1}{c}{$\kappa=-2$}
&\multicolumn{1}{c}{$\kappa=2$}&\multicolumn{1}{c}{$\kappa=-3$}\\ \hline
2.02950&1.94995 &1.95360&2.35035&2.36185 \\
\end{tabular}
\end{ruledtabular}
\end{table}

\subsection{Dynamic multipolar polarizability and hyperpolarizability}
For an atom exposed under a linear polarized laser field with the laser frequency $\omega$, the dynamic magnetic-dipole and electric-quadrupole polarizabilities for the initial state $|0\rangle\equiv|n_0,J_0=0\rangle$ (where $n_0$ represents all other quantum numbers) are written as~\cite{porsev04a}
\begin{eqnarray}
\alpha^{M1}(\omega)&=&\frac{2}{3}\sum_n\frac{\Delta E_{n0}|\langle 0\|M1\|nJ_n\rangle|^2}{\Delta E_{n0}^2-\omega^2}
\,,\label{e1} \\
\alpha^{E2}(\omega)&=&\frac{1}{30}(\alpha\omega)^{2}\sum_n\frac{\Delta E_{n0}|\langle 0\|Q\|nJ_n\rangle|^2}{\Delta E_{n0}^2-\omega^2}
\,,\label{e2}
\end{eqnarray}
where $\alpha$ is the fine structure constant, $M1$ and $Q$ are, respectively, the magnetic-dipole and electric-quadrupole transition operators. $\Delta E_{n0}$ represents the transition energy between the initial state $|0\rangle$ and the intermediate state $|nJ_n\rangle$.

For the $J_0=0$ state, the dynamic hyperpolarizability $\gamma_0(\omega)$ is expressed as,
\begin{eqnarray}
\gamma_0(\omega)=\frac{1}{9}\mathcal{T}(1,0,1,\omega,-\omega,\omega)+\frac{2}{45}\mathcal{T}(1,2,1,\omega,-\omega,\omega)\,, \nonumber \\
\label{e3b}
\end{eqnarray}
with $\mathcal{T}(J_a,J_b,J_c,\omega_1,\omega_2,\omega_3)$ expressed as the followed general formula~\cite{tang14b},
\begin{widetext}
\begin{eqnarray}
\mathcal{T}(J_a,J_b,J_c,\omega_1,\omega_2,\omega_3)=\sum_P\bigg[\sum_{m_am_bm_c}^{\prime}
\frac{\langle 0\|D^{\mu_1}\|m_aJ_a\rangle\langle m_aJ_a\|D^{\mu_2}\|m_bJ_b\rangle\langle m_bJ_b\|D^{\mu_3}\|m_cJ_c\rangle\langle m_cJ_c\|D^{\mu_4}\|0\rangle}
{(\Delta E_{m_a0}-\omega_\sigma)(\Delta E_{m_b0}-\omega_1-\omega_2)(\Delta E_{m_c0}-\omega_1)} \nonumber \\
+(-1)^{J_a+J_c+1}\delta(J_b,J_0)\sum_{m_a}^{\prime}\frac{\langle 0\|D^{\mu_1}\|m_aJ_a\rangle\langle m_aJ_a\|D^{\mu_2}\|0\rangle}{(\Delta E_{m_a0}-\omega_\sigma)}\sum_{m_c}^{\prime}\frac{\langle 0\|D^{\mu_3}\|m_cJ_c\rangle\langle m_cJ_c\|D^{\mu_4}\|0\rangle}{(\Delta E_{m_c0}+\omega_2)(\Delta E_{m_c0}-\omega_1)}\bigg]
\,,\label{e5}
\end{eqnarray}
\end{widetext}
where $D^{\mu_i}$ is the dipole transition operator, and $\omega_i$ are the frequencies of the external electric field in the three directions with
$\omega_\sigma=\omega_1+\omega_2+\omega_3$. $\sum\limits_P$ implies a summation over the 24 terms generated by permuting the pairs$($-$\omega_\sigma/D^{\mu_1})$, $(\omega_1/D^{\mu_2})$, $(\omega_2/D^{\mu_3})$, $(\omega_3/D^{\mu_4})$, the superscripts $\mu_i$ are introduced for the purpose of labeling the permutations~\cite{pipin92a,tang14b}, and the prime over the summation means that the intermediate state of $|m_iJ_i\rangle\equiv|n_0,J_0=0\rangle$ ($i=a,b,c$) should be excluded in Eq.(\ref{e5}).

It's noted that the relationship between our hyperpolarizability $\gamma_0(\omega)$ and the $\beta(\omega)$ of Porsev {\em et al.}~\cite{porsev18a} is $\gamma_0(\omega)=4\beta(\omega)$~\cite{manakov86a}, which indicates both of $\mathcal{T}(1,0,1,\omega,-\omega,\omega)$ and $\mathcal{T}(1,2,1,\omega,-\omega,\omega)$ terms in Eq.(\ref{e3b}) are four times of $Y_{101}(\omega)$ and $Y_{121}(\omega)$ of Ref.~\cite{porsev18a}, respectively. Compared with the dynamic multipolar polarizabilities, the calculation of the dynamic hyperpolarizabilities using the sum-over-states method is much more challenging, since the $\mathcal{T}(J_a,J_b,J_c,\omega_1,\omega_2,\omega_3)$ term involves three summations over a large number of intermediated states. This makes it more difficult to calculate the dynamic hyperpolarizability of the clock atoms with high accuracy.

In present work, we perform a large-scale configuration-interaction calculations by constructing sufficient configurations in an appropriate cavity to make sure the completeness of intermediate states, which guarantees the accuracy of our calculations for the dynamic multipolar polarizabilities and hyperpolarizabilities.

\section{Results and Discussions}
\subsection{Comparisons of energies, reduced matrix elements and static dipole polarizabilities}
In order to test the correctness and reliability of our method, we make detailed comparisons of the energies, reduced matrix elements and static dipole polarizability in Tables~\ref{t2}-\ref{t4} between present results and other available values. From the comparison of the energies in Table~\ref{t2}, the biggest difference between our DFCP+RCI results and NIST energy~\cite{nistasd500} is 0.335\%. From Table~\ref{t3}, it is seen that the difference for all the reduced matrix elements between our results and the values of Ref.~\cite{safronova13b} are within 2\% except the $5s^2\,^1S_0\rightarrow 5s5p\,^3P_1^o$ and $5s^2\,^1S_0\rightarrow 5s6p\,^1P_1^o$ transitions. And from the static electric dipole polarizability in Table~\ref{t4}, we can see that our values are 202.02 a.u. and 465.81 a.u. for the $5s^2\,^1S_0$ and $5s5p\,^3P_0$ clock states respectively, which agree with the results of 198.9 a.u. and 453.4 a.u. of Safronova {\em et al.}~\cite{safronova13b} within 3\%.

Since in the later calculations for the dynamic multipolar polarizabilities and hyperpolarizabilities, we will replace our energies with the NIST energies~\cite{nistasd500}, the error bar of our values mainly comes from reduced matrix elements. From the comparison of the reduced matrix elements in Table~\ref{t3}, although the difference between our values and results of Ref.~\cite{safronova13b} for the $5s^2\,^1S_0\rightarrow 5s5p\,^3P_1^o$ and $5s^2\,^1S_0\rightarrow 5s6p\,^1P_1^o$ transitions is about $-$3.8\% and $-$16.4\%. However, both of them only have about 0.12\% contribution to the ground-state polarizability (see from Table~\ref{t4}), which are much smaller than the 94\% contribution from the $5s^2\,^1S_0\rightarrow 5s5p\,^1P_1^o$ transition. This indicates that the large difference in the reduced matrix elements of $5s^2\,^1S_0\rightarrow 5s5p\,^3P_1^o$ and $5s^2\,^1S_0\rightarrow 5s6p\,^1P_1^o$ transitions between our values and other results has little effect on the final polarizability. Therefore, we can introduce $\pm$3\% fluctuation into all the reduced matrix elements to evaluate conservatively the uncertainty of our multipolar polarizabilities and hyperpolaribilities.
\begin{table}[!htbp]
\caption{\label{t2} Comparison of energy (in cm$^{-1}$) for some selective low-lying states.}
\begin{ruledtabular}
\begin{tabular}{l@{\extracolsep{2em}}l@{\extracolsep{2em}}l@{\extracolsep{2em}}l@{\extracolsep{2em}}l@{\extracolsep{2em}}l@{\extracolsep{2em}}l@{\extracolsep{2em}}l}
\multicolumn{1}{c}{State}&\multicolumn{1}{c}{Present}&\multicolumn{1}{c}{NIST~\cite{nistasd500}}&\multicolumn{1}{c}{Diff.}\\ \hline
$5s^2\,^{1}S_{0}$   &$-$134491.48 & $-$134897.36 &$-$0.301\%\\
$5s6s\,^{1}S_{0}$   &$-$104184.28 & $-$104305.54 &$-$0.116\%\\
$5p^2\,^{3}P_{0}$   &$-$99511.11  & $-$99703.93  &$-$0.193\%\\
$5p^2\,^{1}S_{0}$   &$-$97619.69  & $-$97737.14  &$-$0.120\%\\
$5s7s\,^{1}S_{0}$   &$-$96347.02  & $-$96453.36  &$-$0.110\%\\
$5s8s\,^{1}S_{0}$   &$-$93816.62  & $-$93845.05  &$-$0.030\%\\
$5s9s\,^{1}S_{0}$   &$-$92285.67  & $-$92300.80  &$-$0.016\%\\
$5s10s\,^{1}S_{0}$   &$-$91375.90  & $-$91385.20  &$-$0.010\%\\ \hline
$5s5p\,^{3}P_{0}^o$ &$-$120241.53 & $-$120579.86 & $-$0.281\%\\
$5s6p\,^{3}P_{0}^o$ &$-$100953.10 & $-$101043.88 & $-$0.090\%\\
$4d5p\,^{3}P_{0}^o$ &$-$97533.06  & $-$97605.30  & $-$0.074\%\\
$5s7p\,^{3}P_{0}^o$ &$-$95471.60  & $-$95485.70  & $-$0.015\%\\
$5s8p\,^{3}P_{0}^o$ &$-$93174.39  & $-$93185.32  & $-$0.012\%\\
$5s9p\,^{3}P_{0}^o$ &$-$91904.58  & $-$91911.51  & $-$0.008\%\\ \hline
$5s5p\,^{3}P_{1}^o$ & $-$120066.55 & $-$120393.03 & $-$0.271\% \\
$5s5p\,^{1}P_{1}^o$ &$-$113209.78  & $-$113198.92 & $-$0.010\% \\
$5s6p\,^{3}P_{1}^o$ &$-$100941.97  & $-$101029.05 & $-$0.086\%\\
$5s6p\,^{1}P_{1}^o$ &$-$100754.48  & $-$100798.97 & $-$0.044\%\\
$4d5p\,^{3}D_{1}^o$ &$-$98662.01   & $-$98633.22  & $-$0.029\%\\
$4d5p\,^{3}P_{1}^o$ &$-$97520.99   & $-$97594.64  & $-$0.075\%\\
$5s7p\,^{1}P_{1}^o$ &$-$95944.97   & $-$95990.51  & $-$0.047\%\\ \hline
$5s5p\,^{3}P_{2}^o$ & $-$119696.91  & $-$119998.82 & $-$0.252\%\\
$4d5p\,^{3}F_{2}^o$ & $-$101482.42  & $-$101630.52 & $-$0.146\%\\
$4d5p\,^{1}D_{2}^o$ & $-$101006.28  & $-$101070.47 & $-$0.064\%\\
$5s6p\,^{3}P_{2}^o$ & $-$100843.28  & $-$100924.30 & $-$0.080\%\\
$4d5p\,^{3}D_{2}^o$ & $-$98545.47   & $-$98515.62  & $-$0.030\%\\
$4d5p\,^{3}P_{2}^o$ & $-$97484.48   & $-$97560.78  & $-$0.078\%\\
$5s4f\,^{3}F_{2}^o$ & $-$96142.22   & $-$96146.95  & $-$0.005\%\\ \hline
$5s4d\,^{3}D_{1}$   & $-$116417.74 & $-$116738.33 & $-$0.275\% \\
$5s6s\,^{3}S_{1}$   & $-$105819.31 & $-$105858.60 & $-$0.037\%\\
$5s5d\,^{3}D_{1}$   & $-$99792.66  & $-$99890.46  & $-$0.098\%\\
$5p^{2}\,^{3}P_{1}$   & $-$99310.15  & $-$99497.26  & $-$0.188\%\\
$5s7s\,^{3}S_{1}$   & $-$97457.80  & $-$97472.69  & $-$0.015\% \\
$5s6d\,^{3}D_{1}$   & $-$95168.98  & $-$95211.54  & $-$0.045\%  \\ \hline
$5s4d\,^{3}D_{2}$   & $-$116359.19 & $-$116678.58 & $-$0.274\% \\
$5s4d\,^{1}D_{2}$   & $-$114362.94 & $-$114747.68 & $-$0.335\% \\
$5s5d\,^{1}D_{2}$   & $-$100039.09 & $-$100169.92 & $-$0.131\% \\
$5s5d\,^{3}D_{2}$   & $-$99778.03  & $-$99875.38  & $-$0.097\% \\
$5p^{2}\,^{3}P_{2}$   & $-$99040.99  & $-$99222.73  & $-$0.183\% \\
$5p^{2}\,^{1}D_{2}$   & $-$97769.91  & $-$97936.53  & $-$0.170\% \\
\end{tabular}
\end{ruledtabular}
\end{table}
\begin{table}[!htbp]
\caption{\label{t3} Comparison of some reduced matrix elements (in a.u.), the fourth column is the difference between present values and the results of Ref.~\cite{safronova13b}. }
\begin{ruledtabular}
\begin{tabular}{lllrllllr}
\multicolumn{1}{c}{Transition}&\multicolumn{1}{c}{Present}&\multicolumn{1}{c}{Ref.~\cite{safronova13b}}&\multicolumn{1}{c}{Diff.}&\multicolumn{1}{c}{Others}\\ \hline
$5s^2\,^1S_0\rightarrow 5s5p\,^1P_1^o$  &5.307 &5.272 &0.66\%&5.248(2)$^a$     \\
$5s^2\,^1S_0\rightarrow 5s5p\,^3P_1^o$  &0.152 &0.158 &$-$3.80\%&0.151(2)$^b$ \\
$5s^2\,^1S_0\rightarrow 5s6p\,^1P_1^o$  &0.235 &0.281 &$-$16.4\%&0.26(2)$^c$  \\
$5s5p\,^3P_0^o\rightarrow 5s4d\,^3D_1$  &2.760 &2.712 &1.77\%&2.5(1)$^d$  \\
$5s5p\,^3P_0^o\rightarrow 5s6s\,^3S_1$  &2.002 &1.970 &1.62\%&2.03(6)$^e$   \\
$5s5p\,^3P_0^o\rightarrow 5s5d\,^3D_1$  &2.457 &2.460 &$-$0.12\%&2.3(1)$^f$  \\
$5s5p\,^3P_0^o\rightarrow 5p^2\,^3P_1$  &2.655 &2.619 &1.37\%&2.5(1)$^f$  \\
$5s5p\,^3P_0^o\rightarrow 5s7s\,^3S_1$  &0.523 &0.516 &1.36\%&0.61(2)$^f$  \\
$5s5p\,^3P_0^o\rightarrow 5s6d\,^3D_1$  &1.167 &1.161 &0.52\%&            \\
\end{tabular}
\end{ruledtabular}
\begin{tablenotes}
\item[a]$^a$Ref.~\cite{yasuda06a},$^b$Ref.~\cite{drozdowski97a},$^c$Ref.~\cite{parkinson76a},$^d$Ref.~\cite{miller92a},$^e$Ref.~\cite{jonsson84b},$^f$Ref.~\cite{sansonetti10a}.
\end{tablenotes}
\end{table}
\begin{table}[!htbp]
\caption{\label{t4} Contributions to the static dipole polarizability (in a.u.) of the $5s^2\,^1S_0$ and $5s5p\,^3P_0^o$ states.}
\begin{ruledtabular}
\begin{tabular}{llllllll}
\multicolumn{3}{c}{$5s^2\,^1S_0$}&\multicolumn{3}{c}{$5s5p\,^3P_0^o$}\\
\cline{1-3}\cline{4-6}
\multicolumn{1}{c}{Contr.} &\multicolumn{1}{c}{Present}&\multicolumn{1}{c}{Ref.~\cite{safronova13b}}
&\multicolumn{1}{c}{Contr.} &\multicolumn{1}{c}{Present}&\multicolumn{1}{c}{Ref.~\cite{safronova13b}}\\ \hline
$5s5p\,^1P_1^o$ & 189.947 &187.4  &$5s4d\,^3D_1$ & 290.162 & 280.2\\
$5s5p\,^3P_1^o$ & 0.234   &0.25   &$5s6s\,^3S_1$ & 39.850  & 38.6 \\
$5s6p\,^1P_1^o$ & 0.236   &0.34   &$5s5d\,^3D_1$ & 42.700  & 42.8\\
$4d5p\,^1P_1^o$ & 0.976   &0.95   &$5p^2\,^3P_1$ & 48.932  & 47.6 \\
                &         &       &$5s7s\,^3S_1$ & 1.734   & 1.69 \\
                &         &       &$5s6d\,^3D_1$ & 7.855   & 7.8  \\
Tail            &4.813    & 4.60  &Tail          & 28.766  & 29.1\\
Valance         & 196.206 & 193.54&Valance       & 459.999 & 447.79\\
Core            & 5.812   & 5.29  &Core          & 5.812   & 5.55\\
Total           & 202.02  & 198.9 &Total          & 465.81 & 453.4\\
\end{tabular}
\end{ruledtabular}
\end{table}

\subsection{Comparison of multipolar polarizabilities}
Table~\ref{t5} lists the dynamic magnetic-dipole, electric-quadrupole polarizabilities of the $5s^2\,^1S_0$ and $5s5p\,^3P_0^o$ clock states at the 813.4280(5) nm~\cite{ye08a} magic wavelength. A direct comparison between our work and the calculations of Porsev {\em et al.}~\cite{porsev18a} are also given in this table. The differential $M1$ polarizability $\Delta\alpha^{M1}(\omega)$ is determined thoroughly by $\alpha^{M1}_{^3P_0^o}(\omega)$, since the $\alpha^{M1}_{^3P_0^o}(\omega)$ polarizability is more than 3 orders of magnitude larger than $\alpha^{M1}_{^1S_0}(\omega)$ polarizability. The differential $E2$ polarizability $\Delta\alpha^{E2}(\omega)$ is an order of magnitude larger than $\Delta\alpha^{M1}(\omega)$. The final value of the differential dynamic multipolar polarizability $\Delta\alpha^{QM}(\omega)$ is $2.68(94)\times10^{-5}$ a.u., which agrees well with the CI+All-order result of $2.80(36)\times10^{-5}$ a.u.~\cite{porsev18a}.

\begin{figure}
\includegraphics[width=0.48\textwidth]{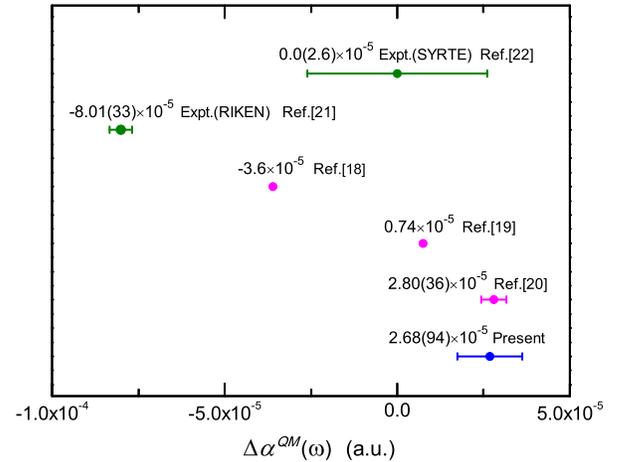}
\caption{\label{f1}(Color online)  Comparison of the $\Delta\alpha^{QM}(\omega)$ (in a.u.). The green line represents measurement results. The blue line represents our present value, and the magenta line denotes other theoretical results. }
\end{figure}

The detailed comparison for the differential multipolar polarizability between theory and experiment are summarized in the Fig.~\ref{f1}. It is seen that for $\Delta\alpha^{QM}(\omega)$, there are obvious differences among values of CI+All-order method~\cite{porsev18a} and the single-electron Fues$^\prime$ model potential (FMP) approach~\cite{katori15a,ovsiannikov13a}. Especially, the recent measurement in RIKEN~\cite{ushijima18a} disagrees with earlier measurement of Ref.~\cite{westergaard11a}, and also disagrees with all the theoretical values, even the sign of $\Delta\alpha^{QM}(\omega)$ between theory and experiment is still opposite. Our work independently validate the CI+All-order calculations of Ref.~\cite{porsev18a}, but differs from the recent experimental measurement~\cite{ushijima18a}. This existing discrepancy deserves further theoretical and experimental investigations on the multipolar polarizabilities of the Sr clock.

\begin{table}[ht]
\caption{\label{t5} The dynamic magnetic-dipole and electric-quadrupole polarizabilities (in a.u.) for the $5s^2\,^1S_0$ and $5s5p\,^3P_0^o$ clock states at the 813.4280(5) nm magic wavelength. $\Delta\alpha^{M1}(\omega)=\alpha^{M1}_{\,^3P_0^o}(\omega)-\alpha^{M1}_{\,^1S_0}(\omega)$ and $\Delta\alpha^{E2}(\omega)=\alpha^{E2}_{\,^3P_0^o}(\omega)-\alpha^{E2}_{\,^1S_0}(\omega)$ represent the difference for the clock states of the dynamic magnetic-dipole and electric-quadrupole polarizabilities, respectively. And $\Delta\alpha^{QM}(\omega)=\Delta\alpha^{M1}(\omega)+\Delta\alpha^{E2}(\omega)$. The numbers in parentheses are computational uncertainties. The numbers in the square brackets denote powers of ten.}
\begin{ruledtabular}
\begin{tabular}{lrrcr}
&\multicolumn{1}{c}{Present}&\multicolumn{2}{c}{Ref.~\cite{porsev18a}}\\
\cline{2-2}\cline{3-4}
\multicolumn{1}{l}{Polarizability}&\multicolumn{1}{c}{DFCP+RCI}&\multicolumn{1}{c}{CI+PT}&\multicolumn{1}{c}{CI+All-order}\\ \hline
  $\alpha^{M1}_{\,^1S_0}(\omega)$   &2.12(13)[-9]     &2.19[-9]     &2.37[-9]      \\
\specialrule{0em}{1pt}{1pt}
  $\alpha^{M1}_{\,^3P_0^o}(\omega)$ &$-$5.05(31)[-6]  &$-$5.09[-6]  &$-$5.08[-6]   \\
\specialrule{0em}{1pt}{1pt}
  $\Delta\alpha^{M1}(\omega)$      &$-$5.05(31)[-6]  &$-$5.09[-6]  &$-$5.08[-6]   \\  \hline
\specialrule{0em}{1pt}{1pt}
  $\alpha^{E2}_{\,^1S_0}(\omega)$   &9.26(56)[-5]     &8.61[-5]     &8.87(26)[-5]  \\
\specialrule{0em}{1pt}{1pt}
  $\alpha^{E2}_{\,^3P_0^o}(\omega)$ &12.44(76)[-5]  &12.1[-5]     &12.2(25)[-5]  \\
\specialrule{0em}{1pt}{1pt}
  $\Delta\alpha^{E2}(\omega)$      &3.18(94)[-5]     &3.50[-5]     &3.31(36)[-5]  \\ \hline
\specialrule{0em}{1pt}{1pt}
  $\Delta\alpha^{QM}(\omega)$      &2.68(94)[-5]     &2.99[-5]     &2.80(36)[-5]  \\
\end{tabular}
\end{ruledtabular}
\end{table}

\subsection{Comparison of hyperpolarizabilities}
The dynamic hyperpolarizabilities of the $5s^2\,^1S_0$ and $5s5p\,^3P_0^o$ clock states at 813.4280(5) nm~\cite{ye08a} magic wavelength for the Sr atom are presented in Table~\ref{t6}. Since there a factor of 4 difference in the definition of the hyperpolarizability between our $\gamma_0(\omega)$ and $\beta(\omega)$ of Porsev~\cite{porsev18a}, we use $\frac{1}{36}\mathcal{T}(1,0,1,\omega,-\omega,\omega)=\frac{1}{9}Y_{101}(\omega)$, and $\frac{1}{90}\mathcal{T}(1,2,1,\omega,-\omega,\omega)=\frac{2}{45}Y_{121}(\omega)$ to make a direct comparison with the calculations of Porsev {\em et al.}~\cite{porsev18a}. Our values for both terms of $\frac{1}{36}\mathcal{T}(1,0,1,\omega,-\omega,\omega)$ and $\frac{1}{90}\mathcal{T}(1,2,1,\omega,-\omega,\omega)$ are much closer to the CI+All-order values than the CI+perturbation theory (PT) results of Ref.~\cite{porsev18a}. The difference of $\frac{1}{36}\mathcal{T}(1,0,1,\omega,-\omega,\omega)$ term for the $5s^2\,^1S_0$ state between present value and CI+All-order~\cite{porsev18a} value is about 8\%. For the $\frac{1}{90}\mathcal{T}(1,2,1,\omega,-\omega,\omega)$ term, the agreement between our values with the results of Ref.~\cite{porsev18a} is much better for the $5s^2\,^1S_0$ state than the $5s5p\,^3P_0^o$ state, that dues to the calculations of dynamic hyperpolarizability for the $5s5p\,^3P_0^o$ state involve much more intermediated states, and the completeness of intermediate states is vital for the reliability of the calculations. Our recommended values of $\frac{1}{4}\gamma_0(\omega)$ are $8.2(2.0)\times 10^5$ a.u. and $-2.01(43)\times 10^7$ a.u. for the $5s^2\,^1S_0$ and $5s5p\,^3P_0^o$ states, respectively, which agree well with the values of Ref.~\cite{porsev18a}. The final recommended value for the differential hyperpoalrizability $\frac{1}{4}\Delta\gamma_0(\omega)$ is $-2.09(43)\times 10^7$ a.u., which is mainly determined by the hyperpolarizability of the $5s5p\,^3P_0^o$ state.

\begin{figure}
\includegraphics[width=0.48\textwidth]{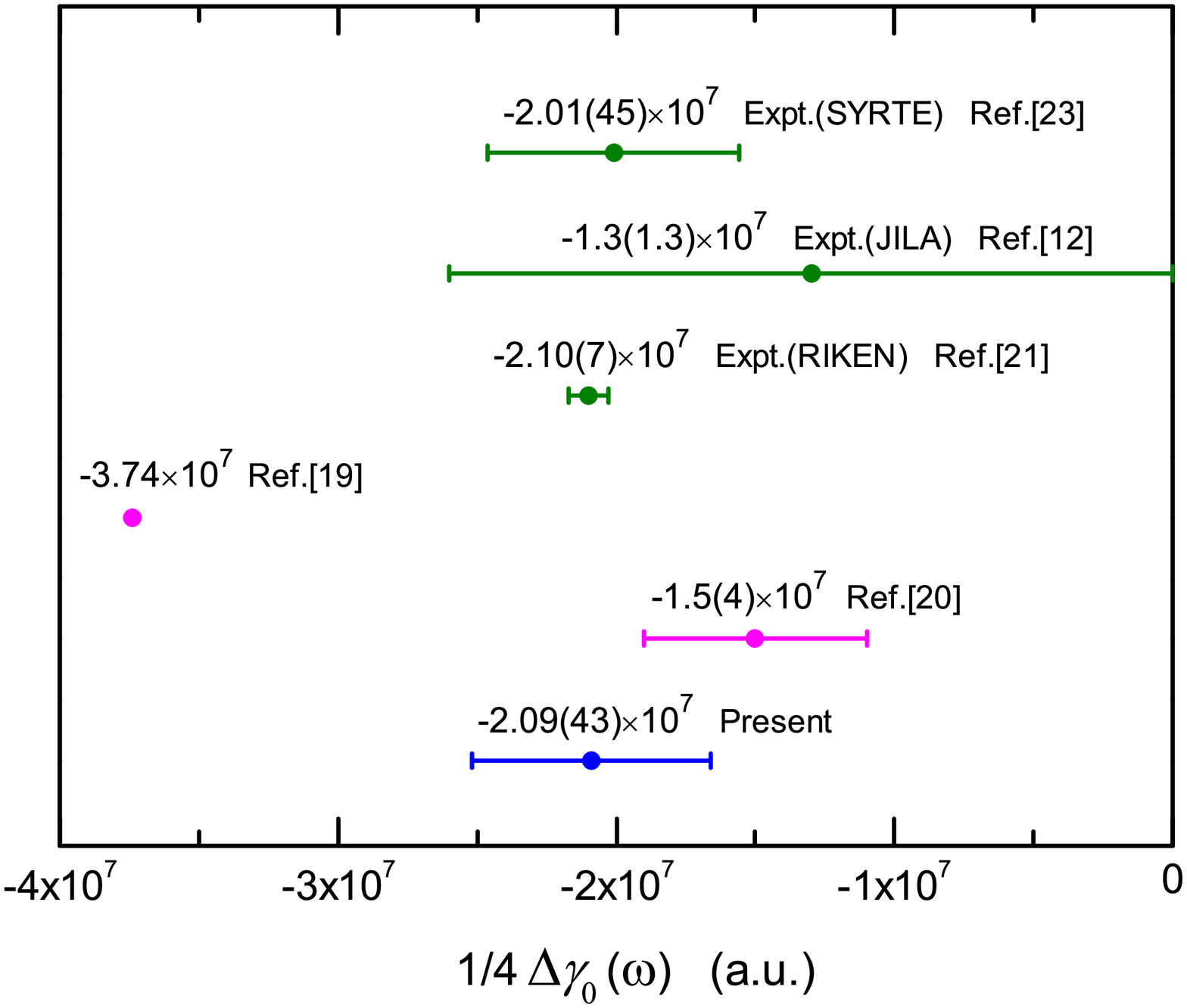}
\caption{\label{f2}(Color online) Comparison of the $\frac{1}{4}\Delta\gamma_0(\omega)$ (in a.u.). The green line represents the experimental values, the blue line represents the present result, and the magenta line denotes other theoretical values.}
\end{figure}

The detailed comparison of the differential hyperpolarizabilitiy for the Sr atom is displayed in Fig.~\ref{f2}. It is seen that the FMP value of $-3.74\times10^{7}$ a.u.~\cite{katori15a} is not within the error bar of any theoretical and experimental results. Two independent theoretical results between our DFCP+RCI value $-2.09(43)\times10^{7}$ a.u. and the CI+All-order result $-1.5(4)\times10^{7}$ a.u. of Ref.~\cite{porsev18a} are both in good agreement with the recent high-accuracy measurement of $-2.10(7)\times10^{7}$ a.u. in RIKEN~\cite{ushijima18a} and $-$2.01(45)$\times10^{7}$ a.u. in SYRTE~\cite{targat13a}.

\begin{table*}[!htbp]
\caption{\label{t6} The dynamic hyperpolarizabilities (in a.u.) for the $5s^2\,^1S_0$ and $5s5p\,^3P_0^o$ clock states at the 813.4280(5) nm magic wavelength. $\frac{1}{36}\mathcal{T}(1,0,1,\omega,-\omega,\omega)=\frac{1}{9}Y_{101}(\omega)$, and $\frac{1}{90}\mathcal{T}(1,2,1,\omega,-\omega,\omega)=\frac{2}{45}Y_{121}(\omega)$, where the definition of $Y_{101}(\omega)$ and $Y_{121}(\omega)$ can refer to the Ref.~\cite{porsev18a}. The numbers in parentheses are computational uncertainties. The numbers in the square brackets denote powers of ten.}
\begin{ruledtabular}
\begin{tabular}{lrrrrrrrrrrrr}
\multicolumn{1}{c}{}&\multicolumn{3}{c}{$5s^2 \,^1S_0$}&\multicolumn{3}{c}{$5s5p \,^3P_0^o$}\\
\cline{2-4}\cline{5-7}
\multicolumn{1}{l}{Hyperpolarizability}&\multicolumn{1}{c}{Present}&\multicolumn{2}{c}{Ref.~\cite{porsev18a}}
&\multicolumn{1}{c}{Present}&\multicolumn{2}{c}{Ref.~\cite{porsev18a}}\\
\cline{2-2}\cline{3-4}\cline{5-5}\cline{6-7}
\multicolumn{1}{c}{}&\multicolumn{1}{c}{DFCP+RCI}&\multicolumn{1}{c}{CI+All-order}&\multicolumn{1}{c}{CI+PT}
&\multicolumn{1}{c}{DFCP+RCI}&\multicolumn{1}{c}{CI+All-order}&\multicolumn{1}{c}{CI+PT}\\ \hline
$\frac{1}{36}\mathcal{T}(1,0,1,\omega,-\omega,\omega)$    &$-$6.70[5]&$-$6.18[5]&$-$6.06[5]&$-$7.88[6]&$-$7.57[6]&$-$7.50[6]\\
\specialrule{0em}{1pt}{1pt}
$\frac{1}{90}\mathcal{T}(1,2,1,\omega,-\omega,\omega)$     &1.49[6]&1.41[6]&1.33[6]&$-$1.22[7]&$-$6.58[6]&$-$3.03[6]\\
\specialrule{0em}{1pt}{1pt}
Total for $\frac{1}{4}\gamma_0(\omega)$ & 8.20[5] & 7.90[5] & 7.25[5]  & $-$2.01[7] & $-$1.42[7] & $-$1.05[7]\\
\specialrule{0em}{1pt}{1pt}
Recommended $\frac{1}{4}\gamma_0(\omega)$ &8.2(2.0)[5]&\multicolumn{2}{c}{7.90(65)[5]} &$-$2.01(43)[7]&\multicolumn{2}{c}{$-$1.42(37)[7]}\\
\specialrule{0em}{1pt}{1pt}
Recommended  $\frac{1}{4}\Delta\gamma_0(\omega)$ &$-$2.09(43)[7]&\multicolumn{2}{c}{$-$1.5(4)[7]}&&&\\
\end{tabular}
\end{ruledtabular}
\end{table*}

\section{Summary}
We have developed the DFCP+CI method for the calculations of the atomic structure properties for divalent atoms. We carried out the calculations of the dynamic magnetic-dipole, electric-quadrupole polarizabilities and hyperpolarizabilities at the magic wavelength for the $5s^2\,^1S_0$ and $5s5p\,^3P_0^o$ clock states of the Sr atom. For the differential hyperpolarizability, our result of $-$2.09(43)$\times$10$^7$ a.u. is in good agreement with the theoretical value of S. G. Porsev {\em et al.}~\cite{porsev18a} and the measurement results of Refs~\cite{westergaard11a,targat13a,ushijima18a}, but disagrees with the zero value measured by JILA~\cite{nicholson15a}. For the differential multipolar polarizability of $\Delta\alpha^{QM}(\omega)$, two independent theoretical calculations from our DFCP+RCI method and the CI+All-order approach of Porsev {\em et al.}~\cite{porsev18a} are consistent with each other, but both have obvious difference from the recent experimental measurement~\cite{ushijima18a}, even the sign of the values is opposite. So the difference about $\Delta\alpha^{QM}(\omega)$ in the Sr clock is still pending, which calls for further experimental investigation to resolve this discrepancy.

\begin{acknowledgments}
We thank Baolong L\"{u}, K. L. Gao, Zhuanxian Xiong and Y. M. Yu for their helpful discussions. We thank S. G. Porsev and M. S. Safronova for their communications. This work was supported by the National Key Research and Development Program of China under Grant No. 2017YFA0304402, by the Strategic Priority Research Program of the Chinese Academy of Sciences under Grants No. XDB21030300, and by the National Natural Science Foundation of China under Grants No. 11774386. Y-B Tang was supported by the National Natural Science Foundation of China No. 11504094.
\end{acknowledgments}


\end{document}